\documentclass[12pt]{article}

\usepackage{graphicx}
\usepackage{epsfig}

\begin{document}
\begin{center}
{\bf Midrapidity hyperon production in pp and pA collisions\\ 
from low to LHC energies} \\
\vspace{0.7cm}

G.H. Arakelyan$^1$, C. Merino$^2$, and Yu.M. Shabelski$^3$ \\

\vspace{.5cm}
$^1$A.Alikhanyan National Scientific Laboratory \\
(Yerevan Physics Institut)\\
Yerevan, 0036, Armenia\\
E-mail: argev@mail.yerphi.am\\
\vspace{0.1cm}

$^2$Departamento de F\'\i sica de Part\'\i culas, Facultade de F\'\i sica\\
and Instituto Galego de F\'\i sica de Altas Enerx\'\i as (IGFAE)\\
Universidade de Santiago de Compostela\\
Santiago de Compostela 15782 \\
Galiza-Spain\\
E-mail: merino@fpaxp1.usc.es \\
\vspace{0.1cm}

$^{3}$Petersburg Nuclear Physics Institute\\
NCR Kurchatov Institute\\
Gatchina, St.Petersburg 188350, Russia\\
E-mail: shabelsk@thd.pnpi.spb.ru
\vskip 0.9 truecm

\vspace{1.2cm}

{\bf Abstract}
\end{center}
 
The experimental data on $p$, $\Lambda$, $\Xi^-$, $\Omega^-$-baryons and the corresponding 
antibaryons spectra obtained by different collaborations are compared with the results
of the calculations performed into the frame of the Quark-Gluon String Model. 
The contribution of String Junction diffusion and the inelastic screening corrections are 
accounted for in the theoretical calculations. The predictions of the 
Quark-Gluon String Model both for $pp$ and $pA$ collisions are extended up to the LHC energies. 

\vskip 1.5cm


\newpage

\section{Introduction}

The study of hadron production has a long history in high-energy particle and nuclear physics,
as well as in cosmic-ray physics. The absolute yields and the transverse momentum ($p_T$) spectra
of identified hadrons in high-energy hadron-hadron collisions are among the most basic
physical observables. They can be used to test the predictions for non-perturbative quantum
chromodynamics (QCD) processes like hadronization and soft-parton interactions, and the validity
of their corresponding implementation in Monte Carlo (MC) event generators~\cite{sjost,scand,deng}.

Among the different available phenomenological models, in this paper we will consider 
the Quark-Gluon String Model (QGSM)~\cite{KTM,K20}, based on the Dual
Topological Unitarization, Regge phenomenology, and nonperturbative
notions of QCD, and that has been successfully used for the description of
multiparticle production processes in hadron-hadron~\cite{KaPi,Sh,AMPS,MPS,ACKS} 
and hadron-nucleus~\cite{KTMS,Sh1,AKMS,pPbold,AMPS2} collisions. 

In the QGSM high energy interactions are considered as proceeding via
the exchange of one or several Pomerons, and all elastic and inelastic
processes result from cutting through or between Pomerons~\cite{AGK}.
Inclusive spectra of hadrons are related to the corresponding fragmentation
functions of quarks and diquarks, which are constructed using the Reggeon
counting rules~\cite{Kai}.

In the case of interaction with a nuclear target, the Multiple Scattering
Theory (Gribov-Glauber Theory) is used, what allows to consider the
interaction with nucleus as the superposition of interactions with
different numbers of target nucleons~\cite{KTMS}.

The QGSM gives a reasonable description of a lot of experimental data on baryons and 
strange hyperons production in hadron-hadron collisions for wide energy range going from 
fixed target experiments up to LHC~\cite{KaPi,Sh,AMPS,ACKS,MPS,KTMS,Sh1,AKMS,pPbold,AMPS2}. 
In this paper we extend the model to the study the multistrange hyperon production.

The multistrange baryons, $\Xi^-$ (dss), and $\Omega^-$ (sss), are particulary important in 
high energy particle and nuclear physics, due to their dominant strange quark (s-quark) 
content. Since the initial-state colliding projectiles contain no strange valence quarks,
all particles in the final state with non-zero strangeness quantum number should have been created
in the course of the collision. Moreover, the energy of the Large Hadron Collider (LHC) and its high
luminosity allow for an abundant production of strange hadrons.
These two factors make multi-strange baryons a valuable probe in 
understanding the particle production mechanisms in high energy collisions.

A remarkable feature of strangeness production is that the production of each additional
strange quark featuring in the secondary baryons, i.e., the production rate of secondary 
$B(qqs)$ over secondary $B(qqq)$, then of $B(qss)$ over $B(qqs)$, and, finally,
of $B(qss)$ over $B(sss)$, is affected by one universal strangeness suppression factor,
$\lambda_s$:
\begin{equation}
\lambda_s = \frac{B(qqs)}{B(qqq)} = \frac{B(qss)}{B(qqs)} = \frac{B(qss)}{B(sss)} \;,
\end{equation}
together with some simple combinatorics~\cite{AKMS,AnSh,CS}.

It will be shown below that $\lambda_s$ slightly increases with initial energy, 
the experimental data favoring the value $\lambda_s$=0.22 at fixed 
target eneries ($\sqrt(s)\le$ 30 GeV), that becomes $\lambda_s$=0.32 at larger 
RHIC and LHC energies. This energy dependence 
can be connected to an increase of hardness of the interacton, i.e. to the increase 
of the secondaries average transverse momenta~\cite{Pi}.

Significant differences in the yields of baryons and antibaryons in the central
(midrapidity) region are evident at not very high energies. This effect can be
naturally explained~\cite{AMPS,MPS,ACKS,Olga,MPS1,MRS} by the structure of baryons
in QGSM, i.e. baryons of three valence quarks together with a specific
configuration of the gluon field, called String Junction~\cite{Artru,IOT,RV,Khar}.

At very high energies the contribution of the enhanced Reggeon diagrams
becomes important, leading to a new phenomenological effect, the suppression
of the inclusive density of secondaries~\cite{CKTr} in the central (midrapidity) 
region.

We present a brief description of the model in Section 2. In Section 3 we 
show the comparison of our numerical calculations with the existing experimental 
data on secondary baryons and antibaryons production in pp and pA collisions.
Conclusion remarks are given in Section 4.

\section{QGSM formalism}
\subsection{General approach}

The QGSM~\cite{KTM,KaPi,Sh} allows one to make
quantitative predictions for different features of multiparticle production,
in particular, for the inclusive densities of different secondaries, both in
the central and in the beam fragmentation regions.

In QGSM, each exchanged Pomeron corresponds to a cylindrical diagram, and thus, 
when cutting one Pomeron, two showers of secondaries are produced (see Fig.~1 a,b).

\begin{figure}[htb]
\vskip -8.cm
\hskip 3.cm
\includegraphics[width=.8\hsize]{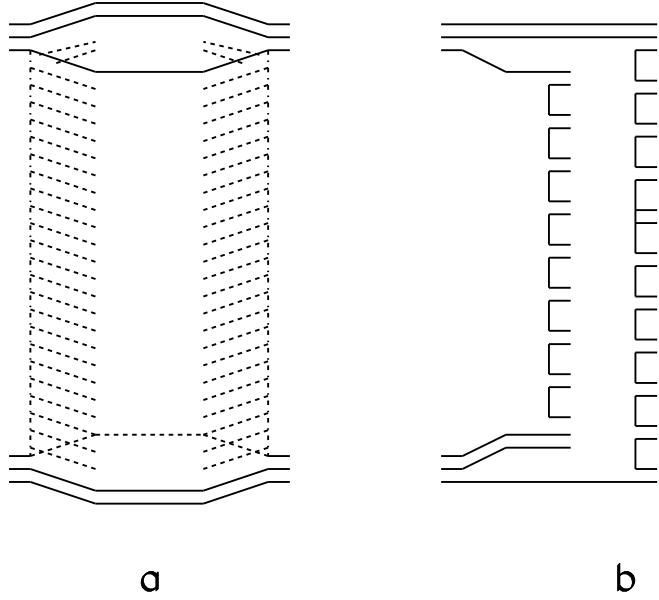}
\vskip -.8cm
\caption{\footnotesize
(a) Cylindrical diagram representing a Pomeron exchange within the DTU
classification (quarks are shown by solid lines); (b) One cut of the cylindrical
diagram corresponding to the single-Pomeron exchange contribution in inelastic
$pp$ scattering.}
\end{figure}

The inclusive spectrum of a secondary hadron $h$ is then determined by the
convolution of the diquark, valence quark, and sea quark distributions,
$u(x,n)$, in the incident particles, with the fragmentation functions, $G^h(z)$,
of quarks and diquarks into the secondary hadron $h$. Here $n$ is the number of 
cutted Pomerons in the considered diagrams. Both the distributions and the fragmentation 
functions are constructed using the Reggeon counting rules~\cite{Kai}.

In particular, in the case of $n > 1$, i.e. in the case of multipomeron
exchange, the distributions of valence quarks and diquarks are softened due to
the appearance of a sea quark contribution. 

The details of the model are presented in~\cite{KTM,KaPi,Sh,ACKS}. The
average number of exchanged Pomerons $\langle n \rangle_{pp}$ slowly
increases with the energy. The values of the Pomeron parameters have been taken
from~\cite{Sh}.

For a nucleon target, the inclusive rapidity, $y$, or Feynman-$x$, $x_F$,
spectrum of a secondary hadron $h$ has the form~\cite{KTM}:
\begin{equation}
\frac{dn}{dy}\
=\frac{x_E}{\sigma_{inel}}\cdot \frac{d\sigma}{dx_F}\ = 
\sum_{n=1}^\infty w_n\cdot\phi_n^h (x) \ ,
\end{equation}
where the functions $\phi_{n}^{h}(x)$ determine the contribution of diagrams
with $n$ cut Pomerons, $w_n$~\cite{KTM} is the relative weight of this diagram.
Here we neglect the numerically small contribution of diffraction dissociation processes.

In the case of $pp$ collisions:
\begin{equation}
\phi_n^{h}(x) = f_{qq}^{h}(x_{+},n) \cdot f_{q}^{h}(x_{-},n) +
f_{q}^{h}(x_{+},n) \cdot f_{qq}^{h}(x_{-},n) +
2(n-1)\cdot f_{s}^{h}(x_{+},n) \cdot f_{s}^{h}(x_{-},n)\ \  ,
\end{equation}

\begin{equation}
x_{\pm} = \frac{1}{2}[\sqrt{4m_{T}^{2}/s+x^{2}}\pm{x}]\ \ ,
\end{equation}
where $f_{qq}$, $f_{q}$, and $f_{s}$ correspond to the contributions
of diquarks, valence quarks, and sea quarks, respectively.

These contributions are determined by the convolution of the diquark and
quark distributions with the fragmentation functions, e.g.,
\begin{equation}
f_{q}^{h}(x_{+},n) = \int_{x_{+}}^{1}
u_{q}(x_{1},n)\cdot G_{q}^{h}(x_{+}/x_{1}) dx_{1}\ \ .
\end{equation}

\subsection{String Junction contribution}

In the string models, baryons are considered as configurations
consisting of three connected strings (related to three valence quarks),
called  String Junction (SJ) \cite{Artru,IOT,RV,Khar}, this picture leading
to some quite general phenomenological predictions.

The production of a baryon-antibaryon pair in the central region usually
occurs via $SJ$-$\overline{SJ}$ pair production (SJ has upper color indices,
whereas anti-SJ ($\overline{SJ}$) has lower indices), which then combines
with sea quarks and sea antiquarks into a $B\overline{B}$ pair~\cite{RV},
as it is shown in Fig.~2a.
\begin{figure}[htb]
\vskip -4.cm
\hskip 2.cm
\includegraphics[width=.7\hsize]{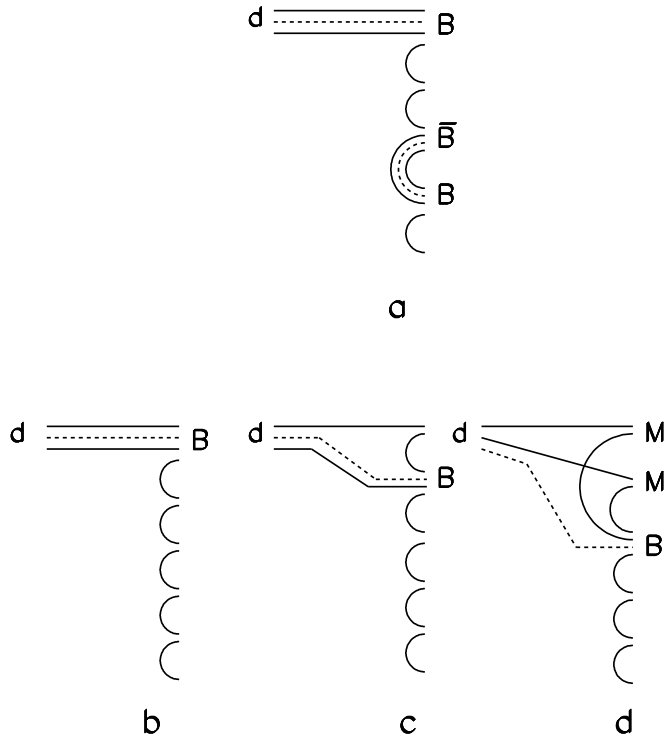}
\caption{\footnotesize
QGSM diagrams describing secondary baryon production:
(a) usual $B\overline{B}$ central production with
production of new SF pair; (b) initial SF together with two valence
quarks and one sea quark; (c) initial SF together with one valence
quark and two sea quarks; (d) initial SF together with three sea quarks.}
\end{figure}

In the case of central production of baryon$-$antibaryon pairs (Fig.~2a) 
the fragmentation functions of any quarks and diquarks into antibaryons are proportional to 
\begin{equation}
\label{ab}
G(z)=a_{\overline{B}}\cdot(1-Z)^{\beta_{\overline{B}}},
\end{equation}
where parameters $a_{\overline{B}}$ and $\beta_{\overline{B}}$
take different values for different antibaryon states.
The parameter $\beta_{\overline{B}}$ was described
in~\cite{ACKS,KTMS,AKMS,BS}.
In the central 
region, $z \rightarrow 0$, the values of the parameter $a_{\overline{B}}$ in the fragmentation
functions of Eq.~5 for $\overline{p}$, $\overline{\Lambda}$, $\overline{\Xi}$, and $\overline{\Omega}$
productions are connected by the following relation~\cite{AKMS}:
\begin{equation}
\label{abr}
a_{\overline{p}} : a_{\overline{\Lambda}} :  a_{\overline{\Xi}} : a_{\overline{\Omega}} =
1 : \sqrt{(5/2)\cdot \lambda_s} : \sqrt{(3/4)\cdot\lambda^2_s}   
 : \sqrt{(1/4)\cdot\lambda^3_s}\ \ ,
\end{equation}
where $\lambda_s$ is the strangeness suppression factor (see Introduction), and the numerical factors 
$\sqrt{5/2}$, $\sqrt{3/4}$, and $\sqrt{1/4}$ are determined by quark
combinatorics~\cite{ACKS,AnSh,CS}. The value of the parameter $a_{\overline{p}}$ for $\overline{p}$ production
was determined by comparison with the experimental data on antiproton production in $pp$ collisions at 
different energies to be~\cite{KTMS,AKMS,BS}
\begin{equation}
\label{aap}
a_{\overline{p}}= 0.18.
\end{equation}

The comparatively large value for $\overline{\Lambda}$ production comes from the fact that many
strange hyperon resonances (e.g. $\Sigma^*(1385)$) decay into $\Lambda$ and not into $\Sigma$. 
 
By using the relations between $a_{\overline{p}}$, $a_{\overline{\Lambda}}$, $a_{\overline{\Xi}}$ and 
$a_{\overline{\Omega}}$ in Eq.~(\ref{abr}),
that derive from simple quark combinatorics~\cite{ACKS,AnSh,CS}, 
we can calculate the $a_{\overline{B}}$ constants in Eq.~(\ref{ab}) corresponding to the central production of
any strange and multistrange antibaryon in the diagrams describing
the central $B\overline{B}$ pair production (see Fig.~2a).  

However, in the processes with incident baryons another possibility exists to produce a secondary
baryon in the central region, called $SJ$ diffusion (see figs.~2b,c,d). 

To obtain the net baryon charge, and according to ref.~\cite{ACKS}, we consider
three different possibilities. The first one is the fragmentation of the
diquark giving rise to a leading baryon (Fig.~2b). A second possibility is
to produce a leading meson in the first break-up of the string and a baryon
in a subsequent break-up~\cite{Kai} (Fig.~2c). In these two first
cases the baryon number transfer is possible only along short distances
in rapidity. In the third case, shown in Fig.~2d, both initial valence
quarks recombine with sea antiquarks into mesons, $M$, while a secondary
baryon is formed by the $SJ$ together with three sea quarks~\cite{MPS,ACKS}.

The fragmentation functions for the secondary baryon $B$ production
corresponding to the three processes shown in Fig.~2b, 2c, and 2d can be
written as follows (see~\cite{ACKS} for more details):
\begin{eqnarray}
G^B_{qq}(z) &=& a_p\cdot v^B_{qq} \cdot z^{2.5} \;, \\
G^B_{qs}(z) &=& a_p\cdot v^B_{qs} \cdot z^2\cdot (1-z) \;, \\
G^B_{ss}(z) &=& a_p\cdot\varepsilon\cdot v^B_{ss} \cdot z^{1 - \alpha_{SJ}}
\cdot (1-z)^2  \;,
\end{eqnarray}
where 
\begin{eqnarray}
a_p = 1.33 
\end{eqnarray}
is the normalization parameter, obtained earlier from the description of proton 
spectra in $pp$ collisions~\cite{KTMS,AKMS,BS}, and $v^B_{qq}$, $v^B_{qs}$, $v^B_{ss}$ are 
the relative probabilities for different baryons production that can be found by simple
quark combinatorics~\cite{AnSh,CS} (see Table~1), and the factor $\varepsilon$ 
accounts for the small probability of the process shown in Fig.~2d to occur with
respect to the processes in Figs.~2b,c. 

\begin{center}
\vskip 0.25cm
\begin{tabular}{|c|c|c|c|c|c|c|}\hline
$B$ & $p$ & $n$ & $\Lambda + \Sigma^0$ & $\Xi^0$ & $\Xi^-$ & $\Omega^-$  \\  
\hline
$v_0$ & $4L^3$ & $4L^3$ & $7.5L^2S$ & $3LS^2$ & $3LS^2$  & $S^3$  \\  
\hline
$v_u$ & $3L^2$ & $L^2$ & $(5/2)LS$ &  $S^2$ & - & -  \\  
\hline
$v_d$ & $L^2$ & $3L^2$ & $(5/2)LS$ & - & $S^2$  & -  \\  
\hline
$v_s$ & - & - & $(5/2)L^2$ & $2LS$  & $2LS$  & $S^3$  \\  
\hline
$v_{uu}$ & $2L$ & - & $(1/4)S$ & - & -  & -  \\  
\hline
$v_{ud}$ & $L$ & $L$ & $S$ & - & - & -  \\  
\hline
$v_{dd}$ & - & $2L$ & $(1/4)S$ & - & - & -  \\  
\hline
$v_{us}$ & - & - & $(5/4)L$ &  $S$ & - & -  \\  
\hline
$v_{ds}$ & - & - & $(5/4)L$ &  - & $S$ & -  \\  
\hline
$v_{ss}$ & - & - & - & $L$ & $L$  & $S$  \\  \hline
\end{tabular}
\vspace{-0.25cm}
\end{center}
{\footnotesize Table 1: The quark combinatorics providing the values of the parameter
$v_i^B$ in eqs.~(8$-$10). Here $S/L= \lambda_s$ and $2L+S=1$.}

The probabilities $v^B_{qq}$, $v^B_{qs}$, $v^B_{ss}$ depend on the vaule of the
strangeness suppression factor $\lambda_s$. In~\cite{ACKS,AKMS} we used 
$\lambda_s$=0.32. 
In the present paper we show that $\lambda_s$ = 0.22 is in a better agreement 
with the main sample of the data at intermediate energies (see section~3).  
 
The fraction $z$ of the incident baryon energy carried by the secondary
baryon decreases from Fig.~2b to Fig.~2d.
Only the processes in Fig.~2d can contribute to the inclusive spectra in the
central region at high energies if the value of the intercept of the $SJ$
exchanged Regge-trajectory, $\alpha_{SJ}$, is large enough. The analysis
in~\cite{MRS} gives a value of $\alpha_{SJ} = 0.5 \pm 0.1$, that is in agreement
with the ALICE Collaboration result, $\alpha_{SJ} \sim 0.5$~\cite{ALICE}. 
In the present calculations we use~\cite{MRS}:
\begin{equation}
\alpha_{SJ}\, =\, 0.5\;\; {\rm and} ~\varepsilon\, =\, 0.0757\, .
\end{equation}

\subsection{\bf Interaction with nuclei at high energies and inelastic screening effects}

In the calculation of the inclusive spectra of secondaries produced in
$pA$ collisions we should consider the possibility of one or several Pomeron
cuts in each of the $\nu$ blobs of proton-nucleon inelastic interactions.
For example, in Fig.~3 it is shown one of the diagrams contributing to the
inelastic interaction of a beam proton with two target nucleons. In the
blob of the proton-nucleon1 interaction one Pomeron is cut, while
in the blob of the proton-nucleon2 interaction two Pomerons are cut.

\begin{figure}[htb]
\vskip -8.cm
\hskip 2.cm
\includegraphics[width=.8\hsize]{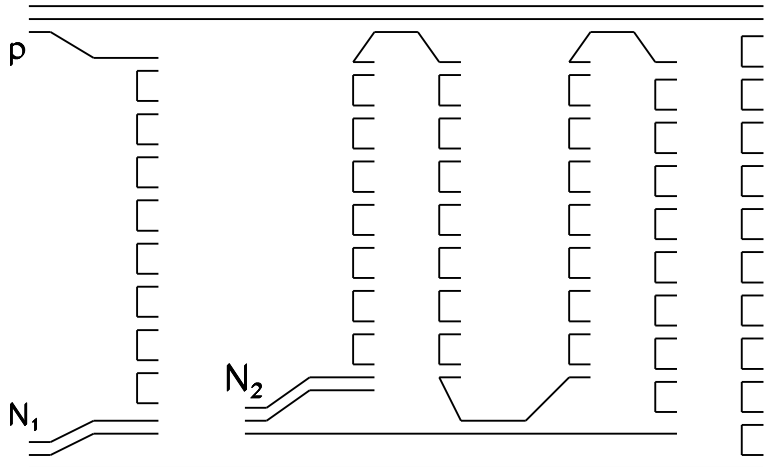}
\vskip -1.6cm
\caption{\footnotesize
One of the diagrams for the inelastic interaction of one
incident proton with two target nucleons $N_1$ and $N_2$ in a $pA$ collision.}
\end{figure}
The contribution of the diagram in Fig.~3 to the inclusive spectrum is:
\vskip -0.7cm
\begin{eqnarray}
\frac{dn}{dy}\ = \
\frac{x_E}{\sigma_{prod}^{pA}}\cdot\frac{d \sigma}{dx_F} & = & 2\cdot
W_{pA}(2)\cdot w^{pN_1}_1\cdot w^{pN_2}_2\cdot\left\{
f^h_{qq}(x_+,3)\cdot f^h_q(x_-,1)\right. + \nonumber\\ \nonumber & + &
f^h_q(x_+,3)\cdot f^h_{qq}(x_-,1) + f^h_s(x_+,3)\cdot[f^h_{qq}(x_-,2) +
f^h_q(x_-,2) + \\ & + & 2\cdot f^h_s(x_-,2)] \left. \right\} \;,
\end{eqnarray}
where $W_{pA}(2)$ is the probability of interaction with namely two target
nucleons.

It is essential to take into account all diagrams with every possible Pomeron
configuration and its corresponding permutations. The diquark and quark distributions and
the fragmentation functions are the same as in the case of $pN$ interaction.

The total number of exchanged Pomerons becomes as large as
\begin{equation}
\langle n \rangle_{pA} \sim
\langle \nu \rangle_{pA} \cdot \langle n \rangle_{pN} \;,
\end{equation}
where $\langle \nu \rangle_{pA}$ is the average number of inelastic
collisions inside the nucleus (about 4 for heavy nuclei at fixed target energies).

The process shown in Fig.~3 satisfies~\cite{Sh3,BT,Weis,Jar} the condition
that the absorptive parts of the hadron-nucleus amplitude are determined by
the combination of the absorptive parts of the hadron-nucleon amplitudes.

The QGSM assumption of the superposition picture of hadron-nucleus interactions 
gives a reasonable description~\cite{MPS,KTMS,Sh4} of the 
inclusive spectra of different secondaries produced at energies $\sqrt{s_{NN}}$ = 14$-$30 GeV.  

At RHIC energies the situation drastically changes. The spectra of
secondaries produced in $pp$ collisions are described rather well~\cite{MPS},
but the RHIC experimental data for $Au+Au$ collisions~\cite{Phob,Phen} give clear
evidence of the inclusive density suppression 
effects which reduce by a factor $\sim$0.5 the midrapidity inclusive density,
when compared to the predictions based on the superposition picture~\cite{CMT,Sh6}.
This reduction can be explained by the inelastic screening
corrections connected to multipomeron interactions~\cite{CKTr} (see Fig.~4). 

\begin{figure}[htb]
\centering
\vskip -8.cm
\hskip 1.5cm
\includegraphics[width=.8\hsize]{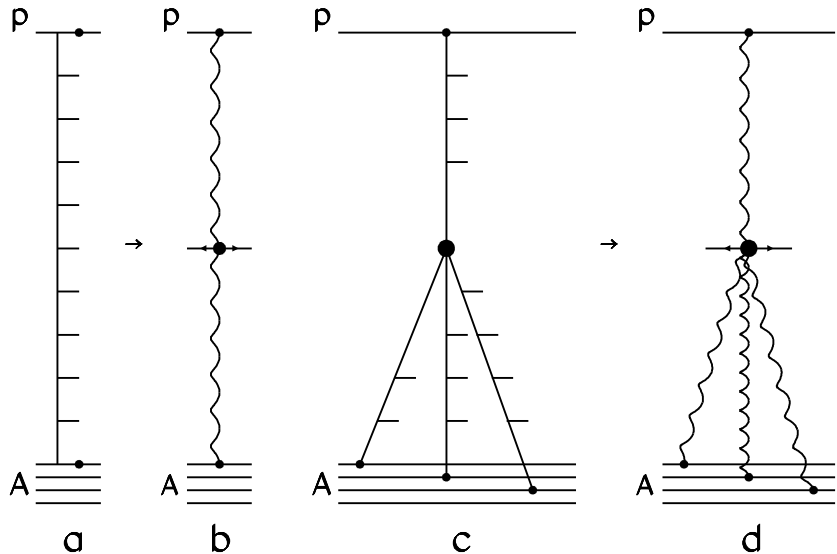}
\vskip -.5cm
\caption{\footnotesize
(a) Multiperipheral ladder corresponding to the inclusive cross section of
diagram (b),  and (c) fusion of several ladders corresponding to the inclusive
cross section of diagram (d).}
\end{figure}

At energies $\sqrt{s_{NN}} \leq $ 30$-$40 GeV, the inelastic processes are 
determined by the production of one (Fig.~4a) or several (Fig.~4c) multiperipheral
ladders, and the corresponding inclusive cross sections are described by the diagrams
of figs.~4b and~4d. 

In accordance with the Parton Model~\cite{Kan,NNN}, the fusion of multiperipheral 
ladders shown in Fig.~4c becomes more and more important with the increase of the
energy, resulting in the reduction of the inclusive density of secondaries. Such
processes correspond to the enhancement Reggeon diagrams of the type of Fig.~4d,
and to even more complicate ones. All these diagrams are proportional to the squared
longitudinal form factor of nucleus $A$~\cite{CKTr}. On the contrary, the 
contribution of such diagrams becomes negligible when the energy decreases,
and they are also negligible for $pp$ collisions up to LHC energies.

In any case, all quantitative estimations of the importance of the contribution
of the enhanced diagrams are model dependent, since the numerical weight of the 
contribution of the multipomeron diagrams is rather unclear due to the many 
unknown vertices in these diagrams. In some models the number of unknown parameters
can be reduced, and, as an example, in reference~\cite{CKTr} the 
Schwimmer model~\cite{Schw} was used for the numerical estimations.

Another possibility to weight the contribution of the
diagrams with Pomeron interaction comes~\cite{JUR,JUR1,BP,JDDSh,BJP}
from Percolation Theory. The percolation approach and its previous version, 
the String Fusion Model~\cite{SFM,SFM1,SFM2}, predicted the multiplicity
suppression seen at RHIC energies, long before any RHIC data were measured.

In order to account for the percolation (inelastic screening) effects in the 
QGSM, it is technically more simple~\cite{MPS1} to consider a maximal number
of Pomerons $n_{max}$ emitted by one nucleon in the central region that 
can be cut. These cut Pomerons lead to the different final states. Then, 
the contributions of all diagrams with $n \leq n_{max}$ are accounted for as 
at lower energies. The larger number of Pomerons $n > n_{max}$ can also be 
emitted obeying the unitarity constraint, but due to the fusion in the final state 
(at the quark-gluon string stage), the cut of $n > n_{max}$ Pomerons results 
in the same final state as the cut of $n_{max}$ Pomerons.

By doing this, all model calculations become very similar to those in the percolation
approach. The QGSM fragmentation formalism allows one to 
calculate the integrated over $p_T$ spectra of different secondaries as
functions of the rapidity and $x_F$.

In this frame, we obtain a reasonable agreement with the experimental data 
on the inclusive spectra of secondaries produced in $d+Au$ collisions 
at RHIC energy~\cite{MPS1} with a value $n_{max} = 13$, and in $p+Pb$ collisions
at LHC energy~\cite{pPbold} with the value $n_{max} = 23$. It has been 
shown in~\cite{JDDCP} that the number of strings that can be used for the secondary
production should increase with the initial energy. 

\section{Numerical results}
\subsection{The pp collisions}

The existing experimental data
for the integrated over the whole range of $p_T$
energy dependence 
of $p$, $\overline{p}$, $\Lambda$, $\overline{\Lambda}$, and $\Xi^-$,
$\overline{\Xi}^+$ hyperons production density $dn/dy (\mid y\mid \leq 0.5)$ 
in $pp$ collisions, from fixed target
to LHC energies~\cite{ATLAS,CMS,NA49,chap,brick,lopinto,jaeger,sheng,kichimi,STAR0} are 
compared with the results of the QGSM calculations in Table~2 and in Fig.~5. 

\begin{figure}[htb]
\centering
\vskip -8.cm
\vskip 2.cm
\includegraphics[width=.8\hsize]{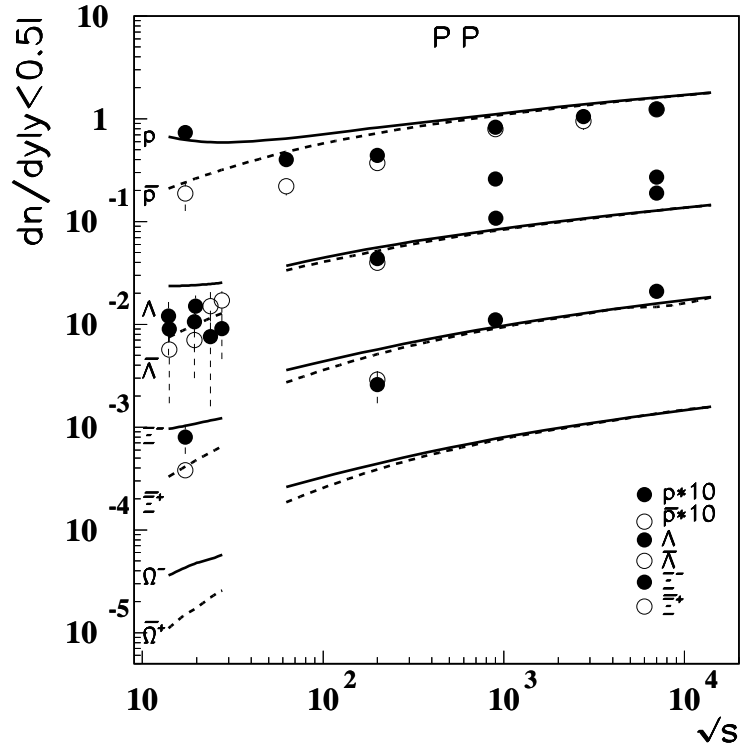}
\vskip -0.8cm
\caption{\footnotesize
The energy dependence of baryon and antibaryon production inclusive densities $dn/dy(\mid~y\mid \leq 0.5)$ 
in the midrapidity region in $pp$ collisions (baryons are shown by closed points and full 
curves, and antibaryons by open points and dashed curves).} 
\end{figure}

The QGSM calculation for secondary $p$ and $\overline{p}$ production in $pp$ collisions
performed with the same values of the parameters $a_{\overline{p}}$=0.18 (see Eq.~7) and
$a_p$=1.33 (see Eq.~11) used in our previous
papers~\cite{KTMS,BS}, are in reasonable agreement with the
experimental data~\cite{NA49,ABZC50,PHENIX,STARp,CMSp,ALICEp09t,ALICEp7t}.

For all baryons ($p$, $\Lambda$, $\Xi$, $\Omega$) and for their corresponding antibaryons, the calculated 
values for dn/dy$(\mid y\mid \leq 0.5)$ are shown in Fig.~5 by full curves, and for antibaryons 
by dashed curves. The theoretical curves and the data on $p$ and $\overline{p}$ production
in Fig.~5 are shown as multiplied by a factor 10. 

The data on $\Lambda$ and $\overline{\Lambda}$~\cite{CMS,ATLAS} productions are reasonably described at 
the lower energies with the value $\lambda_s$=0.22, that should be increased up to $\lambda_s$=0.32 
at $\sqrt{s} \ge $60~GeV. This increase of the value of $\lambda_s$ can be connected to the increase in the
hardness of the interaction, since then the non-zero value of the $s$-quark mass becomes not so significant.
Possibly, the same effect also leads to the increase of average $p_T$~\cite{Olga}.
At intermediate energies the hardness of the interaction is relatively small and the production of the
$s$-quark, with a mass $m_s \sim 0.3 GeV$, is seriously suppressed. At higher enetgies, the hardness of the
interaction increases and the effect of the finite $s$-quark mass becomes not so important, what is refleted
by the effective increase of the value of the $\lambda_s$ parameter.
At extremely high energies, as the hardness of the interaction is much larger,
we expect $\lambda_s \rightarrow$~1. 
 
The yields of $\Xi^-$ and $\overline{\Xi}^+$ produced at 158 GeV/c~\cite{NA49} and at LHC energies
of 0.9 and 7 Tev~\cite{CMS} are well reproduced by correspondingly using the same values of $\lambda_s$=0.22
and $\lambda_s$=0.32. 

In Fig.~5 we also present our prediction for $\Omega^-$ and $\overline{\Omega}^+$ 
hyperon production. 

We do not include in our analysis the ALICE Collaboration data for $\Xi^-$, $\overline{\Xi}^+$, $\Omega^-$
and $\overline{\Omega}^+$ hyperons production~\cite{ALICEms7t} at 7~TeV for the reason that these data were
measured at transverse momenta $p_T > 0.6$ GeV/c, and therefore they can not be compared to integrated over
the whole range of $p_T$ rapidity distributions.
\vskip -1.cm 
{\footnotesize
\begin{center}
\vskip -10pt
\begin{tabular}{|c|c|c|c|c|c|c|} \hline

$\sqrt{s}$ (GeV) & Reaction & QGSM & Experiment dn/dy $(\mid y\mid \leq 0.5)$  \\
\hline

17.29   & p + p $\to p $ & 0.0605 & $0.0736 \pm 0.015$ \cite{NA49p}  \\  

       & p + p $\to \overline{p}$ & 0.024 & $0.054 \pm 0.0014$ \cite{NA49p}  \\  \hline

27.45   & p + p $\to p $ & 0.0576 & $0.0736 \pm 0.015$ \cite{ABZC50}  \\  

        & p + p $\to \overline{p}$ & 0.032 & $0.01869 \pm 0.001$ \cite{ABZC50}  \\  \hline

62.4  & p + p $\to p$ & 0.064 & $0.040 \pm 0.003$ \cite{PHENIX}  \\  

       & p + p $\to \overline{p} $ & 0.0477 & $0.022 \pm 0.002$ \cite{PHENIX}  \\  \hline

200. & p + p $\to p$ & 0.0454 & $0.044 \pm 0.004 $ \cite{PHENIX}  \\  

     & p + p $\to \overline{p}$ & 0.0392 & $0.037 \pm 0.003$ \cite{PHENIX} \\ \hline

900. & p + p $\to p $ & 0.111 & $0.106 \pm 0.005$ \cite{CMSp}  \\  
  & p + p $\to p $ & 0.111 & $0.083 \pm 0.006$ \cite{ALICEp09t}  \\  

  & p + p $\to \overline{p} $ & 0.107 & $0.079 \pm 0.005$ \cite{ALICEp09t}  \\ 

  & p + p $\to \overline{p} $ & 0.107 & $0.104 \pm 0.003$ \cite{CMSp}  \\  \hline

2760. & p + p $\to p $ & 0.138 & $0.124 \pm 0.003$ \cite{CMSp}  \\  

      & p + p $\to \overline{p} $ & 0.135 & $0.1235 \pm 0.0035$ \cite{CMSp}  \\  \hline

7000. & p + p $\to p $ & 0.161 & $0.15 \pm 0.003$ \cite{CMSp}  \\ 
      &                & 0.161 & $0.124 \pm 0.009$ \cite{ALICEp7t}  \\   
      & p + p $\to \overline{p} $ & 0.160 & $0.147 \pm 0.0035$ \cite{CMSp}  \\
      &                    & 0.160 & $0.123 \pm 0.01$ \cite{ALICEp7t}  \\  \hline





17.29  & p + p $\to \Lambda$ & 0.0219 & $0.017 \pm 0.0004$ \cite{NA49}  \\  

       & p + p $\to \overline{\Lambda}$ & 0.0065 & $0.0057 \pm 0.0002$ \cite{NA49}  \\  \hline

%
%

23.79     & p + p $\to \Lambda$ & 0.0227 & $0.0132 \pm 0.003 $ \cite{lopinto}  \\  

            & p + p $\to  \overline{\Lambda}$ & 0.0105 & $0.010 \pm 0.003$ \cite{lopinto}  \\  \hline

27.45     & p + p $\to \Lambda$ &0.0233 & $0.0091 \pm 0.0015$ \cite{kichimi}  \\  

        & p + p $\to \overline{\Lambda}$ & 0.0116 & $0.017 \pm 0.003 $ \cite{kichimi}  \\  \hline

200. & p + p $\to \Lambda$ & 0.0454 & $0.0436 \pm 0.0008 \pm 0.004$ \cite{STAR0}  \\  

      & p + p $\to \overline{\Lambda}$ & 0.0392 & $0.0398 \pm 0.0008 \pm 0.0037$ \cite{STAR0} \\ \hline

%
%
900. & p + p $\to \Lambda$ & 0.0659 & $0.26 \pm 0.01$ \cite{ATLAS} \\  \hline

7000. & p + p $\to \Lambda$ & 0.0987 & $0.27 \pm 0.01$ \cite{ATLAS} \\ \hline

900. & p + p $\to \Lambda$ &0.0659 & $0.108 \pm 0.001 \pm 0.012$ \cite{CMS} \\ \hline

7000. & p + p $\to \Lambda$ & 0.0987& $0.189 \pm 0.001 \pm 0.022$ \cite{CMS} \\  \hline

17.29 & p + p $\to \Xi^- $ & 0.000839 & $0.000799  \pm 0.000095$ \cite{NA49} \\   

     & p + p $\to \overline{\Xi}^+$ & 0.000236 & $0.000381 \pm 0.000057$ \cite{NA49} \\ 
\hline

200. & p + p $\to \Xi^- $ & 0.00542  & $0.0026 \pm 0.0002 \pm 0.0009$ \cite{STAR0} \\   

    & p + p $\to \overline{\Xi}^+$ & 0.00487 & $0.0029 \pm 0.0003 \pm 0.006$ \cite{STAR0} \\ \hline
  
900. & p + p $\to \Xi^-$ & 0.00942 & $0.011 \pm 0.001 \pm 0.001 $ \cite{CMS} \\ \hline

7000. & P + P $\to \Xi^-$ & 0.0159 & $0.021 \pm 0.001 \pm 0.003 $  \cite{CMS} \\   

      & P + P $\to \Xi^-$ & 0.0159 & $0.008 \pm 0.001 \pm 0.007 $  \cite{ALICEms7t} \\
   
     & p + p $\to \overline{\Xi}^+ $ & 0.0158 & $0.0078 \pm 0.001 \pm 0.007$ \cite{ALICEms7t} \\ \hline

200. & p + p $\to \Omega^- + \overline{\Omega^+} $ & 0.0000634 & $0.00034 \pm 0.00016 \pm 0.0005$ \cite{STAR0} \\ \hline

7000. & p + p $\to \Omega^-  $ & 0.00132 & $0.00067 \pm 0.00003 \pm 0.00008$  \cite{ALICEms7t} \\  

       & p + p $\to \overline{\Omega}^+ $ & 0.00134 & $0.00068 \pm 0.00003 \pm 0.00008$ \cite{ALICEms7t} \\ \hline


\end{tabular}
\end{center}

\noindent
Table 2. The experimental data on the inclusive density $dn/dy$ of baryons and antibaryons 
produced in $pp$ collisions at different energies and their corresponding description by the QGSM.}

In Fig.~6 we compare our QGSM calculations with the experimental data
on the rapidity dependence of the production cross section $d\sigma/dy$, integrated over the whole
$p_T$ range, of proton and antiproton production in $pp$ collisions at two energies, 158~GeV/c~\cite{NA49}
and 400~Gev/c~\cite{ABZC50}. The full curve corresponds to the proton production and the
dashed curve to $\overline{p}$ production, both at 158~GeV/c~\cite{NA49}, while the dashed-dotted curve
corresponds to $p$ production and the dotted curve
to $\overline{p}$ production at 400~GeV/c~\cite{ABZC50}.

On the lower panel of Fig.~6, the $\Lambda$ and $\overline{\Lambda}$ production cross section $d\sigma/dy$,
at 158~GeV/c~\cite{NA49} and at 405~GeV/c~\cite{kichimi} in $pp$ collisions are compared with the results of the
QGSM calculations.
The full curve corresponds to $\Lambda$ production at 158~GeV/c and the dashed curve to $\overline{\Lambda}$
production at the same energy, while the dashed-dotted curve corresponds to $\Lambda$ production and the
dotted curve to $\overline{\Lambda}$ production, both at 405~Gev/c.
All curves for $\Lambda$ and $\overline{\Lambda}$ production have been obtained with a value of $\lambda_s$=0.22.
As one can see in Fig.~6, the model calculations agree with the experimetal data.

\begin{figure}[htb]
\vskip -5.cm
\hskip 3.cm
\includegraphics[width=.65\hsize]{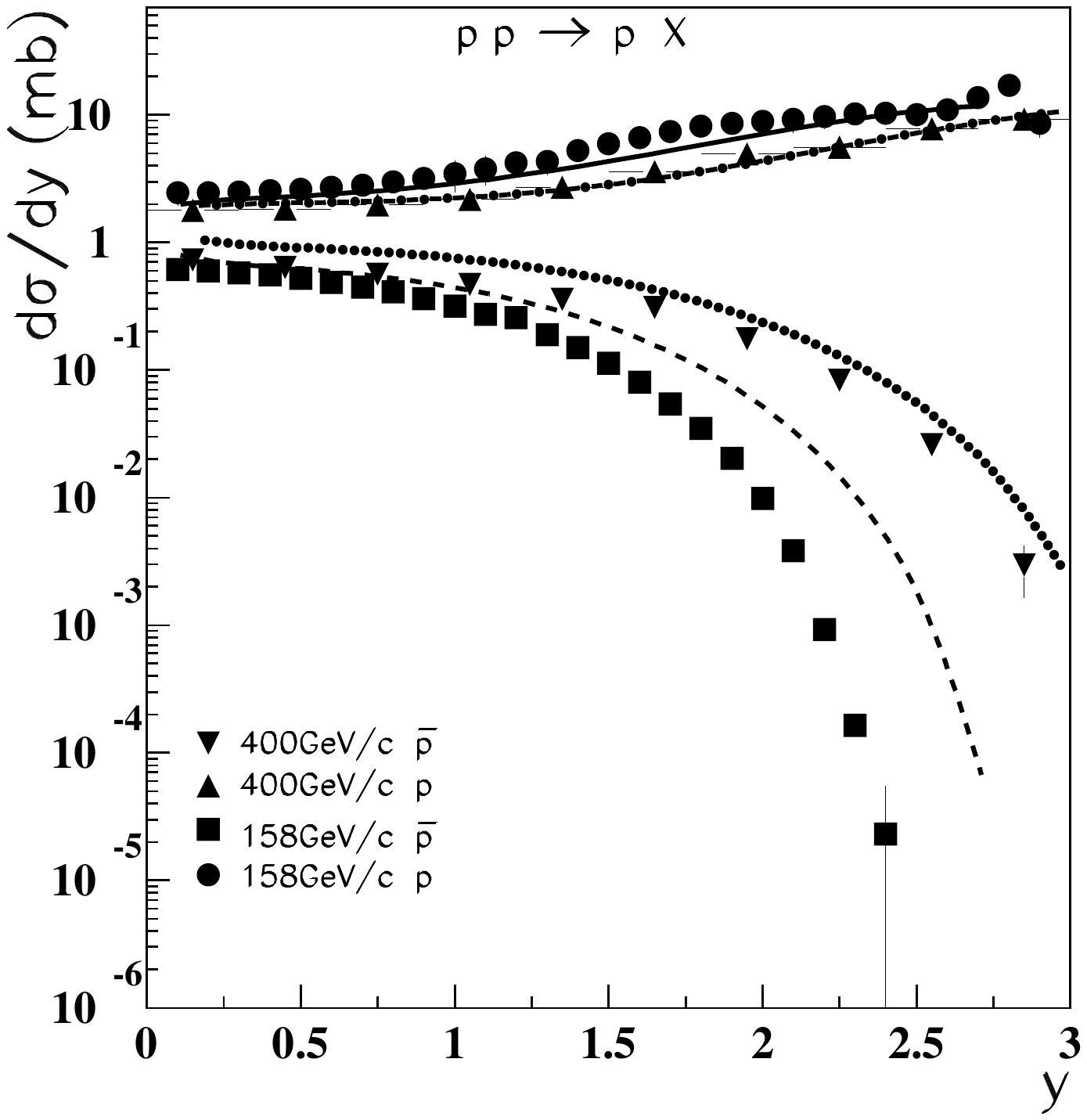}
\vskip -6.cm
\hskip 3.cm
\includegraphics[width=.65\hsize]{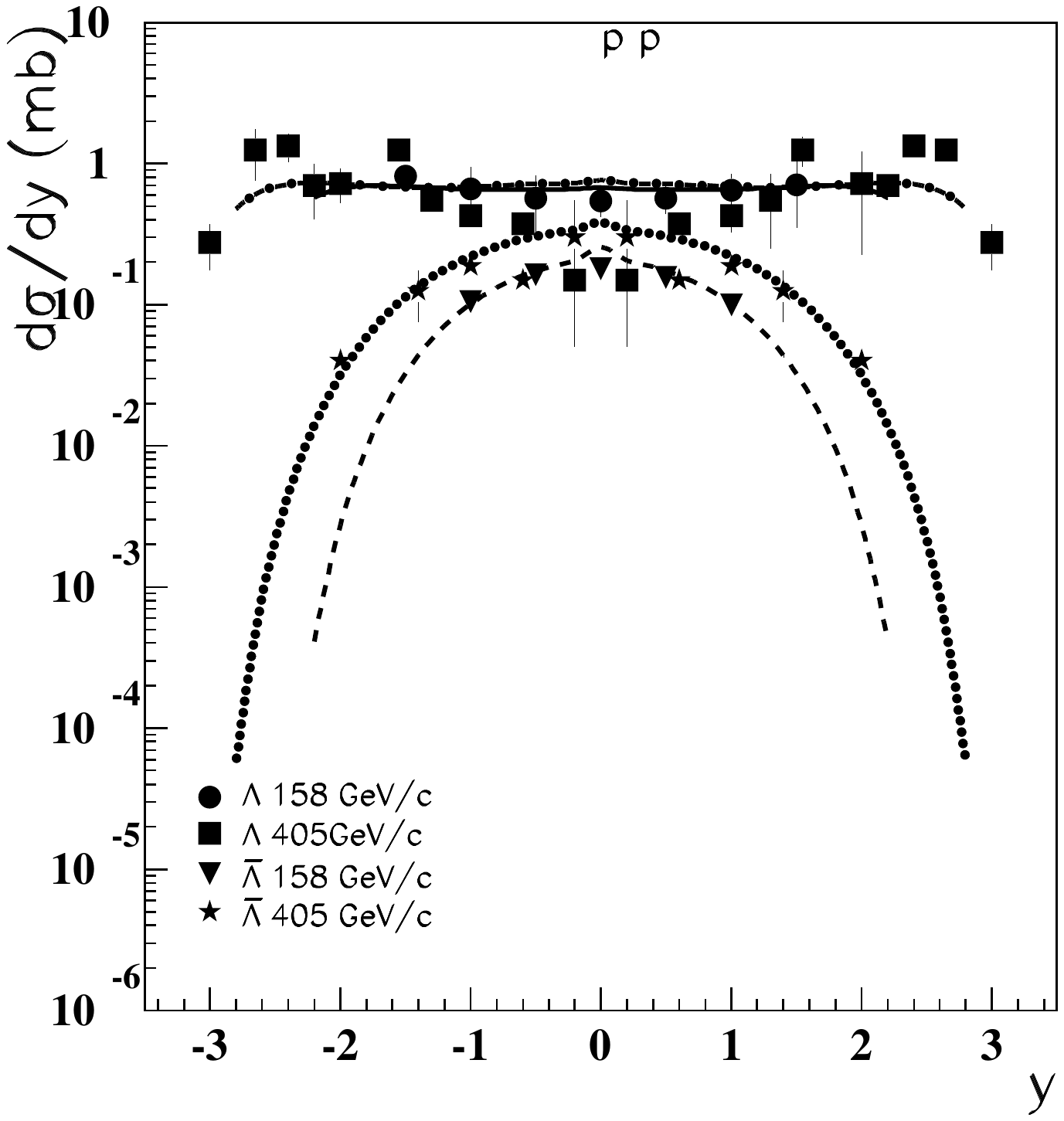}
\vskip -1.cm
\caption{\footnotesize The QGSM results for (upper panel) the rapidity dependence of the
production densities $dn/dy$ of $p$ and $\overline{p}$ at 158~GeV/c
and 400~GeV/c, and their comparison with the experimental data~\cite{NA49,ABZC50},
and for (lower panel) the $\Lambda$ and $\overline{\Lambda}$ cross sections
at 158~GeV/c and 405~GeV/c, and their comparison with the experimental
data~\cite{NA49,kichimi} (see the main text por the description of the different curves).}
\end{figure}

In Fig.~7 we compare the QGSM calculations for the rapidity dependence of the inclusive cross section $d\sigma/dy$
integrated over the whole $p_T$ range of $\Xi^-$ and $\overline{\Xi}^+$ produced in $pp$ collisions
at 158~GeV/c~\cite{NA49}. Also here, the theoretical curves have been data obtained by using the vaue
$\lambda_s$=0.22. The full curve corresponds to $\Xi^-$ production
and the dashed line to the production of $\overline{\Xi}^+$.
 
On the upper panel of Fig.~8 we show the data on the inclusive density $(1/n_{ev})dn/dy$
for $\Lambda$~\cite{ATLAS,CMS} and $\Xi^-$~\cite{CMS} hyperons production at both 0.9 and 7 TeV. 
Here it has to be noticed that the data by ATLAS~\cite{ATLAS} and CMS~\cite {CMS} collaborations
at LHC energies 0.9 and 7~TeV are inconsistent with each other, and, in particular, the ATLAS data 
show a weak energy dependence, while the CMS data would indicate a significantly stronger energy dependence.

As one can see in Fig.~8, the QGSM calculations are in a satisfactory agreement for the 
cases of $\Lambda$ production at 0.9~TeV and for $\Xi^-$ production, both at 0.9 and 7~TeV, 
and they are below the experimental points for $\Lambda$ production at 7~TeV.

\begin{figure}[htb]
\vskip -7.5cm
\hskip 2.cm
\includegraphics[width=.9\hsize]{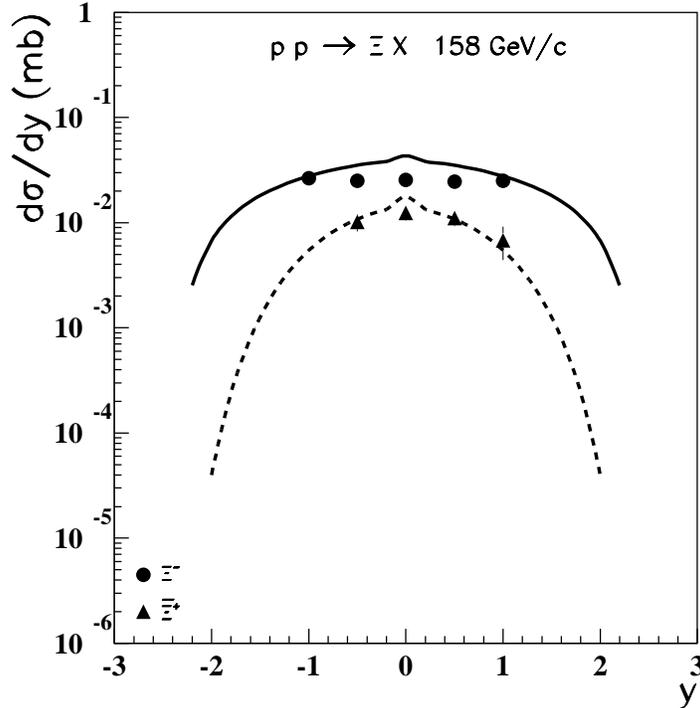}
\vskip -1.5cm
\caption{\footnotesize
The QGSM results for the rapidity dependence of the inclusive cross section $d\sigma/dy$
of $\Xi^-$ and $\overline{\Xi}^+$ productions in $pp$ collisions at 158 GeV/c, and their
comparison with the experimental data~\cite{NA49}. The full curve corresponds to $\Xi^-$
and the dashed curve to $\overline{\Xi}^+$ production.}
\end{figure}

\begin{figure}[htb]
\vskip -6.75cm
\hskip 3.cm
\includegraphics[width=.65\hsize]{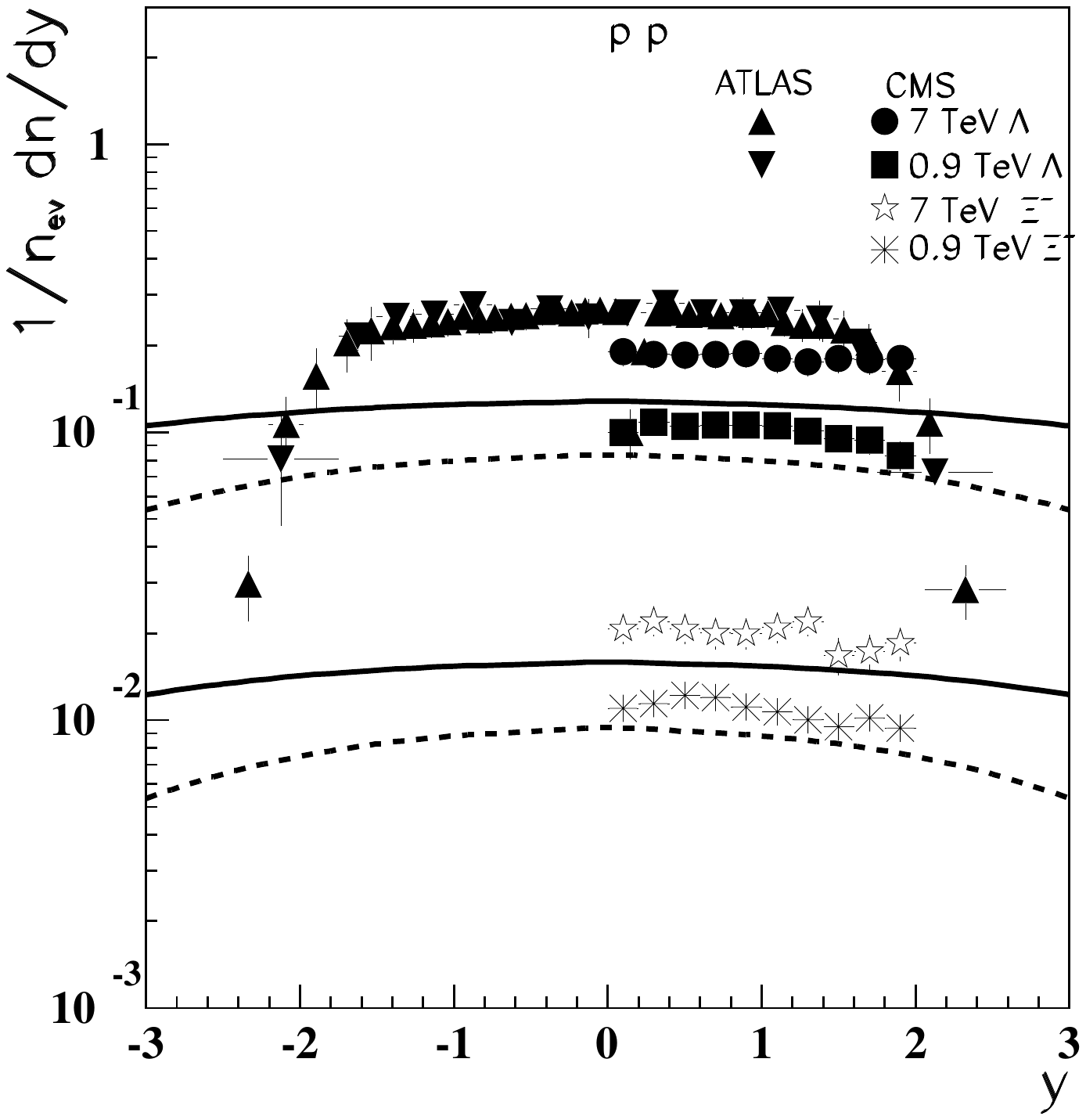}
\vskip -9.75cm
\hskip 2.5cm
\includegraphics[width=.85\hsize]{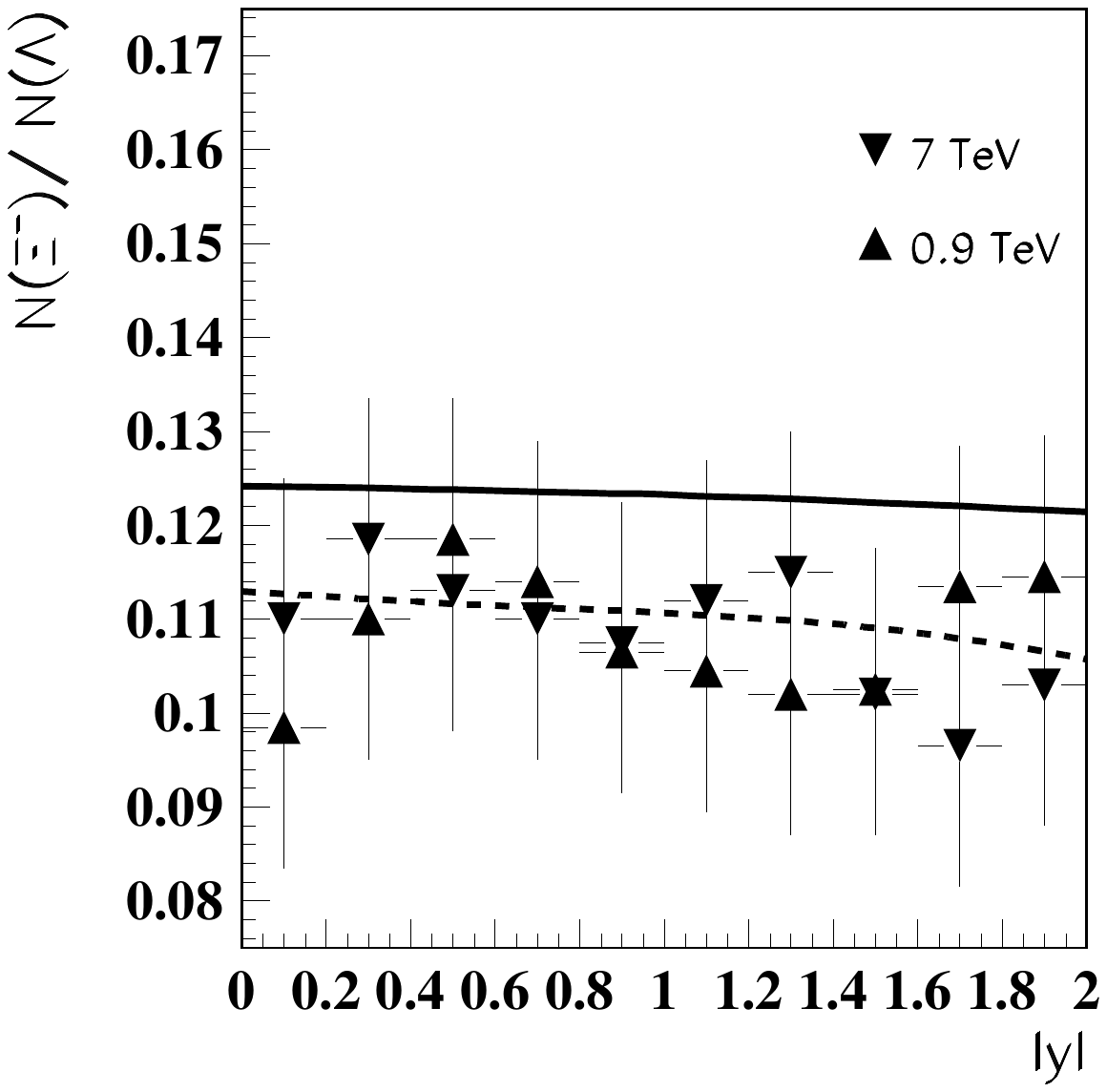}
\vskip -1.5cm
\caption{\footnotesize
The rapidity dependence of (upper panel) the inclusive density $(1/n_{ev})dn/dy$ of $\Lambda$
and $\Xi^-$ hyperons production in the midrapidity region for $pp$ collisions
at LHC energies~\cite{CMS,ATLAS}, and of (lower panel) of the ratio of yields of $\Lambda/\Xi^-$
hyperons at LHC energies~\cite{CMS} in the midrapidity region for $pp$ collisions~\cite{CMS}.
The corresponding QGSM predictions are shown by the full curve ($\sqrt{s}$ = 7~Tev) and by
the dashed curve ($\sqrt{s}$ = 0.9 TeV).}
\end{figure}

On the lower panel of Fig.~8 we show the experimental data on rapidity dependence of ratios of 
$\Lambda/\Xi^-$ densities at 0.9 and 7 TeV~\cite{CMS} and their comparison with the corresponding
QGSM results.
The QGSM reasonably reproduces both the values and the rapidity dependences of the
$\Lambda/\Xi^-$ ratios at the two energies.

\subsection{The $pA$ collisions}

To perform the calculations corresponding to the case of proton-nucleus interactions we have used
the same values of the parameters $a_p$, $a_{\overline{p}}$, and $\lambda_s$ 
as for $pp$ collisions, and we also keep $n_{max}$= 23 (see subsection 2.3) to account for the
inelastic nuclear screening effects. 
\begin{figure}[htb]
\vskip -7.cm
\hskip 2.cm
\includegraphics[width=.9\hsize]{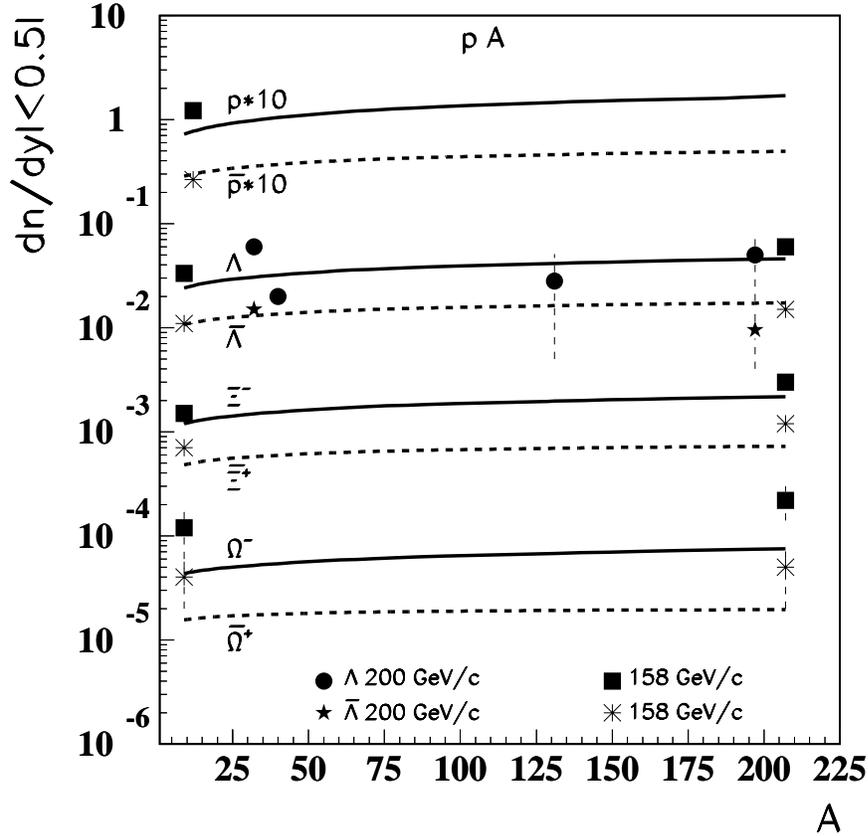}
\vskip -1.cm
\caption{\footnotesize
Comparison of experimental data on the A-dependence of midrapidity density dn/dy $(\mid y\mid \leq 0.5)$ 
of $p$, $\overline{p}$, and $\Lambda$, $\overline{\Lambda}$, $\Xi^-$, $\overline{\Xi}^+$, $\Omega^-$,
and $\overline{\Omega}^+$ hyperons produced at 158 Gev/c and 200 GeV/c~\cite{NA57,NA5,NA35a,NA35b},
with the results of corresponding QGSM calculations (see the main text por the description
of the different curves).}
\end{figure}

In Fig.~9 and Table~3 we compare the results of the QGSM calculations for the A-dependences of $p$, 
$\overline{p}$, and $\Lambda$, $\overline{\Lambda}$, $\Xi^-$, $\overline{\Xi}^+$, $\Omega^-$,
and $\overline{\Omega}^+$ hyperons produced on different nucleus targets with the corresponding
experimental data~\cite{NA57,NA49pC,NA5,NA35a,NA35b}.
The QGSM results at 158~GeV/c are presented for all baryons. For $\Lambda$ and $\overline{\Lambda}$ production
two sets of measurements at 158~Gev/c and at 200~GeV/c exist.
At 158 GeV/c, the full curves correspond to the production of baryons and the dashed curves to
that of antibaryons. The curves corresponding to the calculations at 200 GeV/c are very close to those
obtained at 158 GeV/c, and then not shown.
The QGSM curves are here in a reasonable agreement with the experimental data. 

The experimental data on the rapidity density $dn/dy(\mid y\mid \leq 0.5)$ for $\Lambda$,
$\overline{\Lambda}$, $\Xi^-$, $\overline{\Xi}^+$, $\Omega^-$, and $\overline{\Omega}^+$
hyperons productions at 158~GeV/c and for $p$ and $\overline{p}$ production at
$\sqrt{s}$=5.02~TeV~\cite{CMSpPbp} in $pPb$ collisions~\cite{NA57,NA5,NA35a,NA35b} are presented in Fig.~10,
together with the corresponding QGSM predictions from fixed targets up to LHC energies.

Here, as in Fig.~5, the curves for $p$ and $\overline{p}$ production
and the CMS Collaboration experimental data on average ($p$ + $\overline{p}$)/2 production in $p$+$Pb$ collisions
at $\sqrt{s}$=5.02~TeV~\cite{CMSpPbp}, have been multiplied by a factor 10 to make them distinct from
the $\Lambda$ and $\overline{\Lambda}$ case.

In Fig.~11 we compare the experimental data on the rapidity dependences of the denstity $dn/dy$
of $p$ and $\overline{p}$ produced in $pC$ collisions measured by the NA49 Collaboration at 158 GeV/c~\cite{NA49pC},
together with the corresponding QGSM calculations. The full curve corresponds to $p$ production and the dashed one to
$\overline{p}$ production. The agreement between the QGSM results and the experimental data shown in the figure
is rather good.

{\footnotesize 
\begin{center}
\vskip -10pt
\begin{tabular}{|c|c|c|c|c|c|c|} \hline

$\sqrt{s}$ (GeV) & Reaction   & QGSM & Experiment  dn/dy($\mid y\mid \leq 0.5$)  \\
\hline

17.2 & p + C $\to p $ & 0.0720  & $0.12157 \pm 0.005 $ \cite{NA49pC} \\

    & p + C $\to \overline{p}$ & 0.0285 & $0.02656 \pm 0.0011 $ \cite{NA49pC} \\ 
\hline

5020. & p + Pb $\to (p +\overline{p})/2 $ & 0.440  & $0.5 \pm 0.05 $ \cite{CMSpPbp} \\
\hline

17.2 & p + Be $\to \Lambda$ & 0.0241  & $0.0334 \pm 0.0005 \pm 0.003$ \cite{NA57} \\

   & p + Be $\to \overline{\Lambda}$ & 0.0107 & $0.011 \pm 0.0002 \pm 0.001$ \cite{NA57} \\ 
\hline

19.42 (200 GeV/c) & p + Ar $\to \Lambda$ & 0.0320 & $0.02 \pm 0.01 $ \cite{NA5} \\
\hline

19.42 (200 GeV/c)& P + Xe $\to \Lambda$ & 0.0413 & $0.03 \pm 0.015 $ \cite{NA5} \\
\hline

19.42 (200 GeV/c) & p + Au $\to \Lambda$ & 0.0450 & $0.05 \pm 0.01 $ \cite{NA35a} \\

19.42 (200 GeV/c) & p + Au $\to \overline{\Lambda}$ & 0.0193 & $0.0095 \pm 0.01 $ \cite{NA35a} \\
\hline

19.42 (200 GeV/c) & p + S $\to \Lambda$ & 0.0306 & $0.06 \pm 0.008 $ \cite{NA35b} \\

19.42 (200 GeV/c) & p + S $\to \overline{\Lambda}$ & 0.0131 & $0.015 \pm 0.003 $ \cite{NA35b} \\
\hline

17.2 & p + Pb $\to \Lambda$ & 0.0458 & $0.060 \pm 0.002 \pm 0.006$ \cite{NA57} \\

(m.b.) & p + Pb $\to \overline{\Lambda}$ & 0.0173 & $0.015 \pm 0.001 \pm 0.002$ \cite{NA57} \\ \hline

17.2 & p + Be $\to \Xi^-$ & 0.0011  & $0.0015 \pm 0.0001 \pm 0.0002$  \cite{NA57} \\

(m.b.) & p + Be $\to \overline{\Xi}^+$ & 0.000482 & $0.0007 \pm 0.0001 \pm 0.0002$ \cite{NA57} \\ \hline

17.2 & p + Pb $\to \Xi^-$ & 0.00217  & $0.0030 \pm 0.0002 \pm 0.0003$ \cite{NA57} \\

(m.b.) & p + Pb $\to \overline{\Xi}^+$ & 0.000726 & $0.0012 \pm 0.0001 \pm 0.0001$ \cite{NA57} \\ 
\hline

17.2 & p + Be $ \to \Omega^-$ & 0.0000434 & $0.00012 \pm 0.00006 \pm 0.00002$ \cite{NA57} \\

(m.b.) & p + Be $\to \overline{\Omega}^+$ & 0.0000155 & $0.00004 \pm 0.00002 \pm 0.00001$ \cite{NA57} \\ \hline

17.2 & p + Pb $ \to \Omega^-$ & 0.0000749 & $0.00022 \pm 0.00008 \pm 0.00003$ \cite{NA57} \\

(m.b.) & p + Pb $\to \overline{\Omega}^+$ & 0.0000195 & $0.00005 \pm 0.00003 \pm 0.00002$ \cite{NA57} \\ \hline
\end{tabular}
\end{center}
Table 3. Experimental data for baryons and antibaryons production densities dn/dy $(\mid y\mid \leq 0.5)$
in proton-nucleus collisions at different energies, and the corresponding description by the QGSM.
}
\begin{figure}[htb]
\vskip -7.cm
\hskip 1.5cm
\includegraphics[width=.9\hsize]{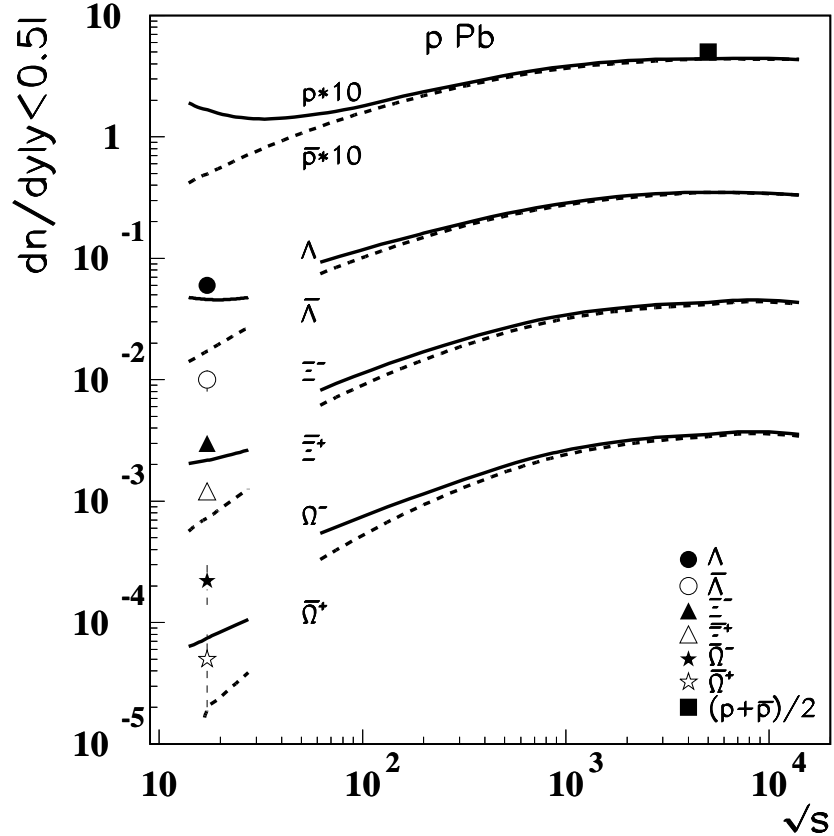}
\vskip -1.cm
\caption{\footnotesize
The energy dependence of the baryon production density $dn/dy(\mid y\mid \leq 0.5)$ in the midrapidity region 
for $pPb$ collisions. The experimental data for average proton production ($p$ + $\overline{p}$/2)
at 5.02~TeV are by the CMS Collaboration~\cite{CMSpPbp}. The experimental data on hyperon production
were measured by the NA57 Collaboration at 158 GeV/c~\cite{NA57}. The curves are the same as in Fig.~4.}
\end{figure}

\begin{figure}[htb]
\vskip -8.75cm
\hskip 1.5cm
\includegraphics[width=1.\hsize]{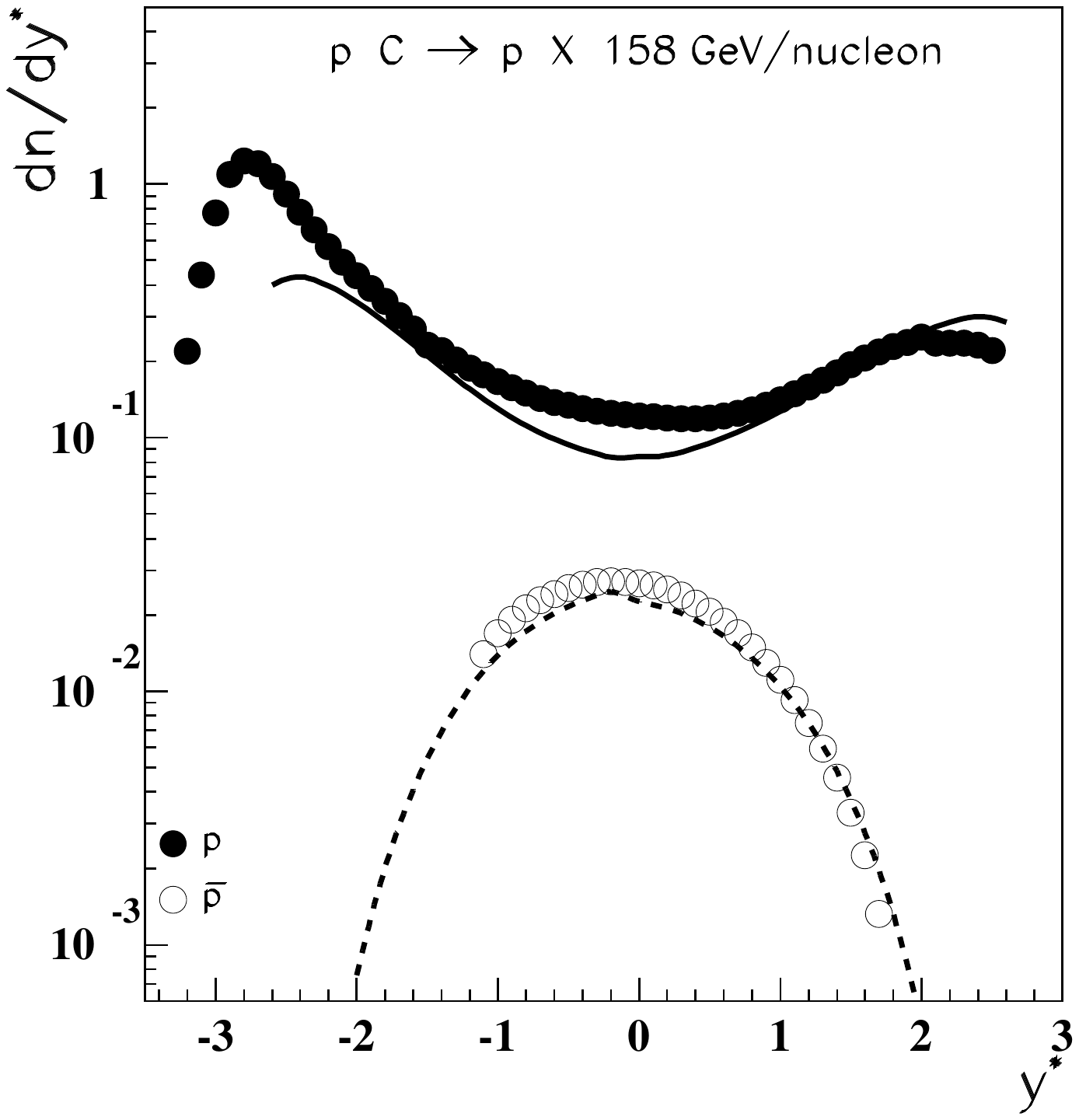}
\vskip -1.25cm
\caption{\footnotesize
The rapidity dependence of the production denstities of $p$ and $\overline{p}$ baryons  
produced in $pC$ collisions measured by the NA49 Collaboration at 158 GeV/c~\cite{NA49pC}, together 
with the results of the QGSM calculations. The theoretical curves correspond to those in Fig~4.}
\end{figure}

In Fig.~12, the rapidity dependence of the production denstity of $\Lambda$ hyperon in proton collisions on $Ar$ and $Xe$ 
nuclei~\cite{NA5} (upper panel) and of $\Lambda$ and $\overline{\Lambda}$ hyperons   
produced in proton collisions on $S$ and $Au$ nuclei~\cite{NA35a,NA35b,NA35c} (lower panel), all 
at 200 GeV/c, are compared with the results of the corresponding QGSM calculations. The full curve on the 
upper panel corresponds to $\Lambda$ production on $Ar$ nucleus and the dash-dotted curve to $\Lambda$ 
production on $Xe$ nucleus. On the lower panel, the full curve corresponds to $\Lambda$ production on 
$Au$ nucleus, and the dash-dotted curve to $\Lambda$ production on $S$ nucleus. Also on the lower panel,
the dashed curve corresponds to $\overline{\Lambda}$ production on $Au$ nucleus and
the dotted curve to $\overline{\Lambda}$ production on $S$ nucleus.
\begin{figure}[htb]
\vskip -7.cm
\hskip 3.cm
\includegraphics[width=.675\hsize]{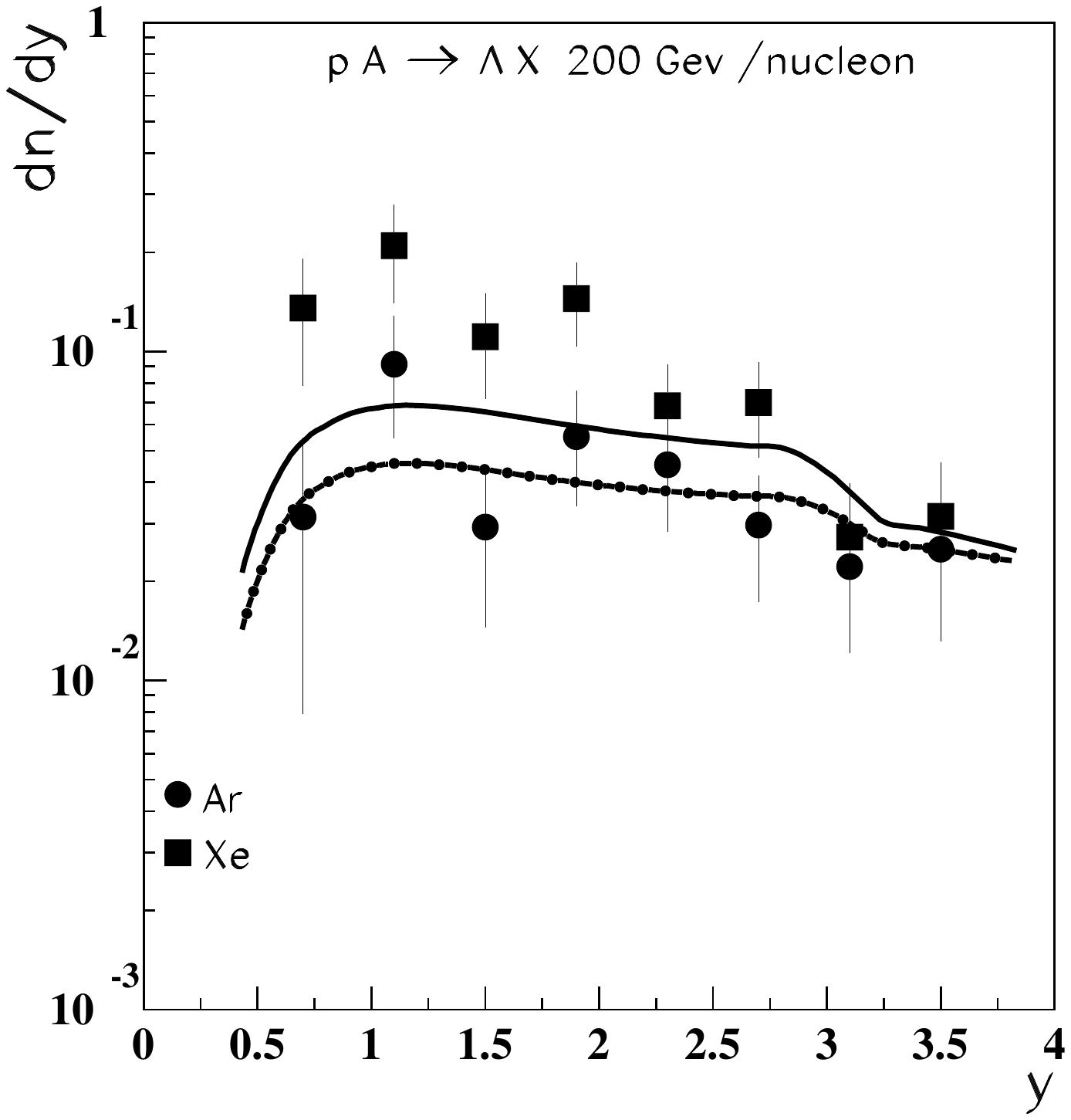}
\vskip -6.2cm
\hskip 3.cm
\includegraphics[width=.675\hsize]{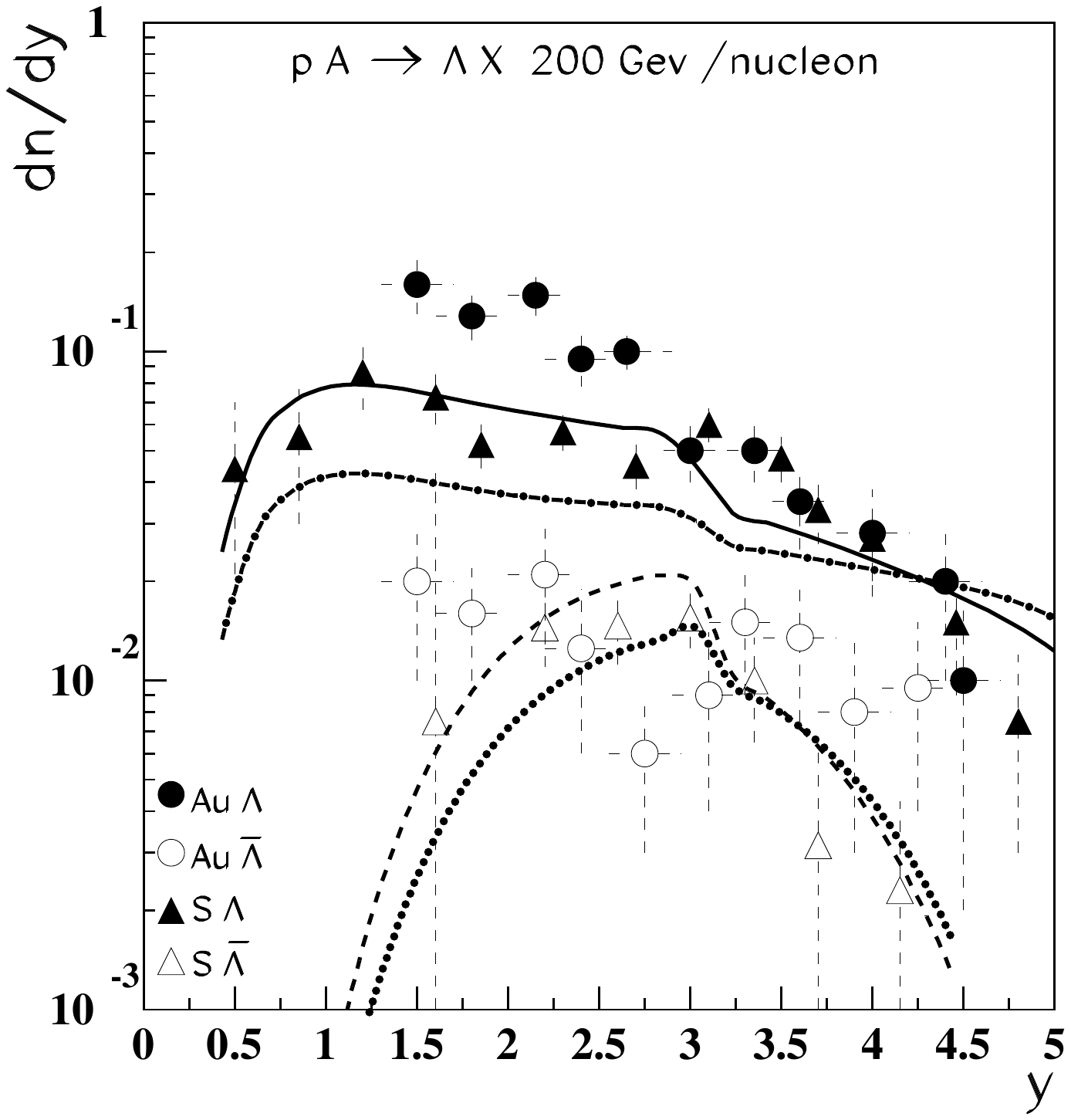}
\vskip -1.15cm
\caption{\footnotesize
The experimental data on the rapidity dependence of the production denstities
of $\Lambda$ hyperon produced in proton collisions on $Ar$ and $Xe$ nuclei~\cite{NA5}
(upper panel) and of $\Lambda$ and $\overline{\Lambda}$ hyperons produced in proton
collisions on $S$ and $Au$ nuclei~\cite{NA35a,NA35b,NA35c} (lower panel), all at 200 GeV/c,
together with the results of the corresponding QGSM calculations
(see the main text por the description of the different curves).}
\end{figure}

One can generally affirm that QGSM results on the rapidity dependence of the production densities
$dn/dy$ of hyperons in $pA$ collisions are in a good agreement with the experimental data at 200~GeV/c,
as it is shown in figs.~11 and 12.

\newpage
\section{Conclusion}

The QGSM, based on the Dual Topological Unitarization, Regge phenomenology, and nonperturbative 
notions of QCD, provides a reasonable description of strange and multistrange hyperons production, 
as well as of the production of their corresponding antiparticles, both in pp and pA collisions 
for a wide range of energies, by including into the analysis the contribution of String 
Junction diffusion and the inelastic screening corrections.

It is important to note that in the QGSM the extension of the calculations of the hyperon (antihyperon)
productions from pp collisions to proton-nucleus collisions does not require any additional
model parameters. 
The accuracy of our calculations can be estimated to be on the level of $\sim 15\%$.

The values of the parameters $a_p$ and $a_{\overline{p}}$ in the model obtained from previous calculations
at lower energies are now confirmed by the good description of p and $\overline{p}$ production at higher
energies. The estimation of the production of all hyperons and antihyperons is governed by the value of
only one parameter $\lambda_s$, which is the same for every unit of strangeness,
but that it depends on energy from the value $\lambda_s$=0.22 at fixed target energies, up to $\lambda_s$=0.32
at LHC energies. In the present paper we do not present any parametrisation of the energy dependence of the value
of the parameter $\lambda_s(E)$, since amount of experimental data is not enough yet to check up this behaviour.

{\bf Acknowledgements}
We thank N.~Armesto and C.~Pajares for useful discussions. We also are grateful to N.I.~Novikova for technical help.

\noindent
This paper was supported by Ministerio de Econom{\'i}a y Competitividad of Spain (FPA2014$-$58293), the Spanish
Consolider-Ingenio 2010 Programme CPAN (CSD2007-00042), by Xunta de Galicia, Spain (2011/PC043), and, in part,
by Russian grant RSGSS-3628.2008.2 and by State Committee of Science of the Republic of Armenia, Grant-13-1C023.-00281.

\end{document}